\documentclass[aps,prd,amsfonts,groupedaddress,showpacs,showkeys]{revtex4}
\usepackage{epsfig}
\newcommand{\eq}{\begin{eqnarray}}
\newcommand{\en}{\end{eqnarray}}
\newcommand{\ra}{\rangle}

\begin{document}

\title{$D^\ast K$ molecular structure of the $D_{s1}(2460)$ meson}
\author{
Amand Faessler,
Thomas Gutsche,
Valery E. Lyubovitskij
\footnote{On leave of absence from the
Department of Physics, Tomsk State University,
634050 Tomsk, Russia},
Yong-Liang Ma
\vspace*{1.2\baselineskip}} 
\affiliation{Institut f\"ur Theoretische Physik,
Universit\"at T\"ubingen,
\\ Auf der Morgenstelle 14, D-72076 T\"ubingen, Germany
\vspace*{0.3\baselineskip}\\}

\date{\today}

\begin{abstract} 
We discuss a possible interpretation of the $D_{s1}(2460)$ 
meson  as a hadronic molecule - a bound state of $D^\ast$ 
and $K$ mesons. Using a phenomenological Lagrangian approach 
we determine the strong $D_{s1} \to D_s^\ast \pi^0$ and 
radiative $D_{s1} \to D_s \gamma$ decays. 
In order of magnitude our results for the partial strong and 
radiative decay widths are consistent with previous calculations. 
\end{abstract}


\maketitle

\newpage

\section{Introduction}

Nowadays there is strong interest to study newly observed mesons 
and baryons in the context of a hadronic molecule interpretation 
(for overview see e.g. Ref.~\cite{Rosner:2006vc}). 
As stressed for example in Ref.~\cite{Barnes:2003dj} the scalar 
$D_{s0}^\ast(2317)$ 
and axial $D_{s1}(2460)$ mesons could be candidates for a scalar $DK$ 
and a axial $D^\ast K$ molecule because of a relatively small binding 
energy of $\sim 50$ MeV. 
These states were discovered and confirmed just a few years ago by 
the Collaborations BABAR at SLAC~\cite{Aubert:2003fg}, CLEO 
at CESR~\cite{Besson:2003cp} and Belle
at KEKB~\cite{Abe:2003jk}. In the interpretation of these experiments it
was suggested that the $D_{s0}^\ast(2317)$ and $D_{s1}(2460)$ mesons are
the $P$-wave charm-strange quark states with spin-parity quantum numbers 
$J^P = 0^+$ and $J^P = 1^+$, respectively. It is worth noting that the 
existing experimental information on the properties of the $D_{s0}^\ast(2317)$ 
and $D_{s1}(2460)$ mesons~\cite{Yao:2006px} leaves quite a significant 
uncertainty in their possible assignment of $J^P = 0^+$ and $J^P = 1^+$ 
quark-antiquark states. 

Strong and radiative decays of the $D_{s0}^\ast(2317)$ and $D_{s1}(2460)$  
mesons have been considered using different 
approaches~\cite{Colangelo:2004vu,Cheng:2003kg,Bardeen:2003kt,%
Godfrey:2003kg,Colangelo:2003vg,Fayyazuddin:2003dp,Ishida:2003gu,%
Azimov:2004xk,Hayashigaki:2004st,Colangelo:2005hv,Close:2005se,%
Wei:2005ag,Liu:2006jx,Lu:2006ry,Guo:2006fu,Guo:2006rp,Wang:2006mf,% 
Faessler:2007gv,Gamermann:2007bm,Gamermann:2007er,Lutz:2007er,%
Lyubovitskij:2007er}, including their different interpretations as 
two- and four-quark states and as $D^{(\ast)}K$ molecules. 
The range of predictions for the strong and radiative decay widths is 
from several to a few hundreds keV. 
In our previous paper~\cite{Faessler:2007gv} 
we calculated the strong $D_{s0}^{\ast} \to D_s \pi^0$ and radiative 
$D_{s0}^{\ast} \to D_s^{\ast} \gamma$ decays using a phenomenological 
Lagrangian approach~\cite{Faessler:2007gv,Faessler:2007cu} for 
the treatment of the $D_{s0}^{\ast}$ meson as a hadronic molecule - 
a bound state 
of $D$ and $K$ mesons. A new feature related to the molecular $DK$
structure of the $D_{s0}^\ast(2317)$ meson was that the presence of $u(d)$
quarks in the $D$ and $K$ mesons gives rise to a direct strong
isospin-violating transition $D_{s0}^{\ast} \to D_s \pi^0$ in
addition to the decay mechanism induced by $\eta-\pi^0$ mixing
as considered previously. We showed that the direct transition
dominates over the $\eta-\pi^0$ mixing transition in the
$D_{s0}^{\ast} \to D_s \pi^0$ decay. 

In this paper we extend our formalism~\cite{Faessler:2007gv,Faessler:2007cu} 
to the strong and radiative decays of the $D_{s1}(2460)$ assuming that the 
latter is a $D^\ast K$ bound state. 
As for the case of the $D_{s0}^{\ast}$ state, a composite (molecular)  
structure of the $D_{s1}(2460)$ meson is defined by the compositeness
condition $Z=0$~\cite{Weinberg:1962hj,Efimov:1993ei,Anikin:1995cf} 
(see also Refs.~\cite{Faessler:2007gv,Faessler:2007cu}). This condition 
implies that the renormalization constant of the hadron wave function 
is set equal to zero or that the hadron exists as a bound state of its 
constituents. The compositeness condition was originally 
applied to the study of the deuteron as a bound state of proton and
neutron~\cite{Weinberg:1962hj}. Then it was extensively used
in low-energy hadron phenomenology as the master equation for the
treatment of mesons and baryons as bound states of light and heavy
constituent quarks (see e.g. Refs.~\cite{Efimov:1993ei,Anikin:1995cf}). 
Constructing the effective mesonic Lagrangian 
including $D_{s1}$, $D^{(\ast)}$, $K^{(\ast)}$ and 
$D^{(\ast)}_s$ degrees of freedom we calculate one-loop meson 
diagrams describing the strong $D_{s1} \to D_s^\ast \pi^0$ and 
radiative $D_{s1} \to D_s \gamma$ decays. A study of two other possible 
radiative decay modes of the $D_{s1}$ meson, the transitions 
$D_{s1} \to D_s^\ast \gamma$ 
and $D_{s1} \to D_{s0}^\ast \gamma$, and, also, the extension 
to the bottom sector ($B_{s0}^\ast(5725)$ and $B_{s1}(5778)$ states) 
will be done in a forthcoming paper. 

In the present manuscript we proceed as follows. First, in Section~II
we discuss the basic notions of our approach. We discuss the effective
mesonic Lagrangian for the treatment of the $D_{s1}(2460)$ meson 
as a $D^\ast K$ bound state. In Section~III we consider the matrix
elements (Feynman diagrams) describing the strong and radiative decays of
the $D_{s1}(2460)$ meson. We discuss our numerical results and perform 
a comparison with other theoretical approaches. 
In Section~IV we present a short summary of our results.

\section{Approach}

\subsection{Molecular structure of the $D_{s1}^{\pm}(2460)$ meson}

In this section we discuss the formalism for the study of the
$D_{s1}^{\pm}(2460)$ meson as a hadronic molecule, represented by 
a $D^\ast K$  bound state. 
We adopt that the isospin, spin and parity quantum numbers of
the $D_{s1}^{\pm}(2460)$ are $I(J^P) = 0(1^+)$, while for its mass 
we take the value $m_{D_{s1}} = 2.4589$ GeV~\cite{Yao:2006px}. 
Our framework is based on an effective interaction Lagrangian describing 
the couplings of the $D_{s1}(2460)$ meson to its constituents:
\eq\label{Lagr_Ds1}
{\cal L}_{D_{s1}}(x) = g_{_{D_{s1}}} \,
D_{s1}^{\mu \, -}(x) \, \int\! dy \,
\Phi_{D_{s1}}(y^2) \, D^{\ast}_\mu(x+w_{_{KD^\ast}} y) \,
K(x-w_{_{D^\ast K}} y) \, + \, {\rm H.c.} \,,
\en
where the doublets of $D^\ast$ and $K$ mesons are defined as 
\eq
D^\ast =
\left(
\begin{array}{c}
D^{\ast \, 0} \\
D^{\ast \, +} \\
\end{array}
\right)\,, \hspace*{1cm}
K =
\left(
\begin{array}{c}
K^+ \\
K^0 \\
\end{array}
\right)\,.
\en
The summation over isospin indices is understood. 
The molecular structure of the $D_{s1}^\pm$ states is  
(we do not consider isospin mixing): 
$|D_{s1}^+\ra \, = \, |D^{\ast +} K^0\ra + |D^{\ast 0} K^+\ra \,,
\hspace*{.2cm}
|D_{s1}^-\ra \, = \, |D^{\ast -} \bar K^0\ra + |\bar D^{\ast 0} K^-\ra \,.$ 
The correlation function $\Phi_{D_{s1}}$ characterizes the finite size 
of the $D_{s1}(2460)$ meson as a $D^\ast K$ bound state and depends on 
the relative Jacobi coordinate $y$ with, in addition, $x$ being the 
center-of-mass (CM) coordinate. In Eq.~(\ref{Lagr_Ds1}) we introduced 
the kinematical parameters $w_{ij} = m_i/(m_i + m_j)$, 
where $m_{D^\ast}$ and $m_K$ are the masses of the $D^\ast$ and $K$ mesons. 
The Fourier transform of the correlation function reads 
\eq
\Phi_{D_{s1}}(y^2) \, = \, \int\!\frac{d^4p}{(2\pi)^4}  \,
e^{-ip y} \, \widetilde\Phi_{D_{s1}}(-p^2) \,. 
\en
A basic requirement for the choice of an explicit form of the correlation
function is that it vanishes sufficiently fast in the ultraviolet region
of Euclidean space to render the Feynman diagrams ultraviolet finite.
We adopt the Gaussian form,
$\tilde\Phi_{D_{s1}}(p_E^2)
\doteq \exp( - p_E^2/\Lambda_{D_{s1}}^2)\,,$
for the vertex function, where $p_{E}$ is the
Euclidean Jacobi momentum. Here, $\Lambda_{D_{s1}}$
is a size parameter, which characterizes the distribution of
$D^\ast$ and $K$ mesons inside the $D_{s1}$ molecule.

The coupling constant $g_{D_{s1}}$ is determined by the compositeness 
condition~\cite{Weinberg:1962hj,Efimov:1993ei,Anikin:1995cf} 
(for an application to $D_{s0}^\ast(2317)$ and $D_{s1}(2460)$ meson 
properties see Refs.~\cite{Faessler:2007gv,Faessler:2007cu}.)  
It implies that the renormalization constant of the hadron
wave function is set equal to zero:
\eq\label{ZDs1}
Z_{D_{s1}} &=& 1 -
\Sigma^\prime_{D_{s1}}(m_{D_{s1}}^2) = 0 \,.
\en
Here, $\Sigma^\prime_{D_{s1}}(m_{D_{s1}}^2)
= g_{_{D_{s1}}}^2 \Pi^\prime_{D_{s1}}(m_{D_{s1}}^2)$ is the
derivative of the transverse part of the mass operator
$\Sigma^{\mu\nu}_{D_{s1}}$, conventionally split into the transverse
$\Sigma_{D_{s1}}$ and longitudinal $\Sigma^L_{D_{s1}}$  parts as:
\eq
\Sigma^{\mu\nu}_{D_{s1}}(p) =
g^{\mu\nu}_\perp \Sigma_{D_{s1}}(p^2) + \frac{p^\mu p^\nu}{p^2}
\Sigma^L_{D_{s1}}(p^2) \,,
\en
where 
$g^{\mu\nu}_\perp = g^{\mu\nu} - p^\mu p^\nu/p^2\,,
\hspace*{.2cm} g^{\mu\nu}_\perp p_\mu = 0\,.$ 
The mass operator of the $D_{s1}$ meson is described by 
the diagram of Fig.1. 

To clarify the physical meaning of the compositeness condition, 
we first want to remind the reader that the renormalization constant 
$Z_{D_{s1}}^{1/2}$ can also be interpreted as the matrix element 
between the physical and the corresponding bare state. 
For $Z_{D_{s1}}=0$ it then follows that the physical state
does not contain the bare one and hence is described as a bound state.
As a result of the interaction of the $D_{s1}$ meson with its constituents 
($D^\ast, K$ mesons), the $D_{s1}$ meson is dressed, i.e. its mass 
and its wave function have to be renormalized. The condition $Z_{D_{s1}}=0$ 
also effectively excludes the tree-level diagrams involving $D_{s1}$ mesons, 
because each external leg of the $D_{s1}$ meson 
is multiplied by the factor~$Z_{D_{s1}}^{1/2}$. 

Following Eq.~(\ref{ZDs1}) the coupling constant $g_{_{D_{s1}}}$
can be expressed in the form: 
\eq\label{gDs1_coupling}
\frac{1}{g_{_{D_{s1}}}^2} = \frac{2}{(4 \pi \Lambda_{D_{s1}})^2} \,
\int\limits_0^1 dx \int\limits_0^\infty
\frac{d\alpha \, \alpha \, P(\alpha, x)}{(1 + \alpha)^3}
\,\, \biggl[ \frac{1}{2 \mu_{D^\ast}^2 (1 + \alpha)}
- \frac{d}{dz} \biggr] \tilde \Phi^2_{D_{s1}}(z)
\label{coupling_Ds1}
\en
where
\eq 
P(\alpha, x)
= \alpha^2 x(1-x) + w_{_{D^\ast \! K}}^2 \alpha x
+ w_{_{KD^\ast}}^2 \alpha (1-x) \,, \ \ \ \ 
z = \mu_{D^\ast}^2 \alpha x + \mu_K^2 \alpha (1-x)
   - \frac{P(\alpha, x)}{1 + \alpha}  \, \mu_{D_{s1}}^2 \,, \ \ \ \
\mu_M = \frac{m_M}{\Lambda_{D_{s1}}} \,. 
\en
Above expressions are valid for any functional form of the correlation
function $\tilde \Phi_{D_{s1}}(z)$.

\subsection{Effective Lagrangian for strong and radiative
decays of $D_{s1}^{\pm}(2460)$}

In the preceding paper~\cite{Faessler:2007gv} in the analysis of 
the strong decay $D_{s0}^\ast \to D_s \pi^0$ we considered the so-called 
``direct'' diagrams (see Fig.2 in \cite{Faessler:2007gv}) 
with $\pi^0$-meson emission from the $D \to D^\ast$ and 
$K \to K^\ast$ transitions and the ``indirect'' diagrams 
(see Fig.3 in \cite{Faessler:2007gv}) where a $\pi^0$ meson 
is produced via $\eta-\pi^0$ mixing in the mass term of pseudoscalar 
mesons in the leading-order $O(p^2)$ Lagrangian of chiral perturbation 
theory (ChPT)~\cite{Gasser:1984gg,Cho:1994zu}. 
Note, that the second mechanism based on $\eta-\pi^0$ mixing
was mainly considered before in the literature. Originally,
it was initiated by the analysis based on the use of chiral
Lagrangians~\cite{Cho:1994zu,Bardeen:2003kt,Colangelo:2003vg}
where the leading-order, tree-level $D_{s0}^\ast D_s \pi^0$ coupling 
can be generated only by virtual $\eta$-meson emission.
During the last years different approaches have been applied to the
$D_{s0}^{\ast} \to D_s \pi^0$ and $D_{s1} \to D_s^\ast \pi^0$ 
decay properties using the $\eta-\pi^0$ mixing mechanism. 
In our approach the $D_{s0}^\ast$ and $D_{s1}$ mesons are considered 
as $DK$ and $D^\ast K$ bound states, respectively and, therefore, we have
an additional mechanism for generating the $D_{s0}^\ast \to D_s \pi^0$ and 
$D_{s1} \to D_s^\ast \pi^0$ transition due to the direct coupling of 
$D^{(\ast)}$ and $K^{(\ast)}$ mesons to $\pi^0$. One should stress, that 
the two types of diagrams (``direct'' and ``mixing'') can be reduced 
to modified pion-emission diagrams performing the diagonalization 
of the pseudoscalar meson mass term~\cite{Gasser:1984gg}. In particular, 
after the diagonalization of the mesonic mass term the $\pi^0$ and $\eta$ 
meson fields are modified by a unitary transformation 
as~\cite{Gasser:1984gg}:  
\eq\label{Unit_trans} 
\pi^0 \ \to \ \pi^0 \cos\varepsilon - \eta \sin\varepsilon \,, 
\hspace*{1cm}
\eta  \ \to \ \pi^0 \sin\varepsilon + \eta \cos\varepsilon  
\en  
with $\varepsilon$ being the $\pi^0-\eta$ mixing angle fixed 
as~\cite{Gasser:1984gg}:  
\eq  
\tan 2\varepsilon = \frac{\sqrt{3}}{2} \frac{m_d - m_u}{m_s - \hat{m}} 
\simeq 0.02 
\,, 
\hspace*{1cm} \hat{m} = \frac{1}{2} (m_u + m_d)
\en
where $m_u, m_d, m_s$ are the current quark masses. 
As a result of the unitary transformation (\ref{Unit_trans}) 
the ``direct'' and ``mixing'' diagrams are combined together in 
the form of pure pion-coupling diagram with modified flavor 
structure, i.e. instead of the $\tau_3 \, \pi^0$ coupling to 
$DD^\ast$ or $KK^\ast$ mesonic pair we have 
$\pi^0 \, ( \tau_3 \, \cos\varepsilon \, + \, \kappa \, I \, 
\sin\varepsilon)$, where $\kappa = 1/\sqrt{3}$ or $\sqrt{3}$ is 
the corresponding flavor-algebra factor for $DD^\ast$ or $KK^\ast$ 
coupling, respectively. 
Below we display the explicit form of the corresponding interaction 
Lagrangian. The lowest-order diagrams which
contribute to the matrix elements of the strong isospin-violating 
decay $D_{s1} \to D_s^\ast \pi^0$ are shown in Fig.2. 
Note, that in the isospin limit ($m_u = m_d$), the $\eta-\pi^0$ 
mixing angle vanishes and the masses of the virtual $D^{(\ast)}$ and 
$K^{(\ast)}$ mesons in the loops are degenerate, respectively. 
As result the pairs of diagrams related to 
Fig.2(a), 2(b) and Fig.2(c) and 2(d) compensate each other. Therefore, 
in the calculation of the diagrams of Fig.2 we go beyond the isospin limit 
and use the physical meson masses. 

The diagrams contributing to the radiative decay
$D_{s1}^{+} \to D_s^{+} \gamma$ are shown in Fig.3.
The diagrams of Figs.3(a) and 3(b) are generated by the direct coupling
of the charged $D^{\ast +}$ and $K^+$ mesons to the electromagnetic field
after gauging the free Lagrangians related to these mesons.
The diagrams of Figs.3(c) and 3(d) (so-called contact diagrams) are generated
after gauging the nonlocal strong Lagrangian~(\ref{Lagr_Ds1})
describing the coupling of $D_{s1}$ meson to its constituents -
$D^\ast$ and $K$ mesons. The diagrams of Figs.3(e) and 3(f) arise after gauging
the strong $D_s D^\ast K$ interaction Lagrangian containing
derivatives acting on the pseudoscalar fields. Finally, the diagrams of
Figs.3(g) and 3(h) describe the subprocess, where the $D_{s1}$ first 
interacts with the electromagnetic field and then converts
into the $D_s$ via a $D^\ast K$ loop. 
Note that an analogous diagram where the $D_{s1}$ converts into 
the $D_s$ and then interacts with the electromagnetic field 
vanishes due to the Lorentz condition for the on-shell
axial meson $D_{s1}$, i.e. $p_\mu \, \epsilon_{D_{s1}}^\mu(p) = 0$.
Details of how to generate the effective couplings of the involved
mesons to the electromagnetic field will be discussed later.
 
After the preliminary discussion of the relevant diagrams,
now we are in the position to write down the full effective Lagrangian
for the study of the strong and radiative decays of the $D_{s1}$ meson 
formulated in terms of mesonic degrees of freedom and photons. 
We follow the procedure discussed 
in detail in Ref.~\cite{Faessler:2007gv}, where we considered 
the $D_{s0}^\ast$ meson decay properties. First, we write the Lagrangian 
${\cal L}$, which includes the free mesonic parts ${\cal L}_{\rm free}$ 
and the strong interaction parts ${\cal L}_{\rm str}$: 
\eq\label{L_full} 
{\cal L}(x) \, = \,
{\cal L}_{\rm free}(x) \, + \, {\cal L}_{\rm str}(x) \,,
\en 
where 
\eq\label{L_free} 
{\cal L}_{\rm free}(x) &=& 
- \frac{1}{2} \vec\pi(x) ( \Box + m_\pi^2 ) \vec{\pi}(x) 
+ \frac{\delta_\pi}{2} \, [\pi^0(x)]^2 
+ D^{+}_{s1, \mu} (x) ( g^{\mu\nu} [\Box + m_{D_{s1}}^2] 
- \partial^\mu\partial^\nu ) D^{-}_{s1, \nu}(x) \nonumber\\
&-&\sum\limits_{P = K, D, D_s} P^\dag(x) ( \Box + m_P^2 ) P(x) 
+ \sum\limits_{P = K, D} \delta_P \, \bar P^0(x) P^0(x) \nonumber\\
&+&\sum\limits_{V = K^\ast, D^\ast, D_s^\ast}  
V^{\dagger}_\mu (x) ( g^{\mu\nu} [\Box + m_{V}^2] 
- \partial^\mu\partial^\nu ) V_\nu(x) 
- \sum\limits_{V = K^\ast, D^\ast} \delta_{V} \, \bar  V^0_\mu (x) 
V^{0 \, \mu} (x) \,, 
\en 
\eq\label{L_strong}   
{\cal L}_{\rm str}(x) & = & - \frac{g_{_{D^\ast D \pi}}}{2\sqrt{2}} 
\, D^{\ast \, \dagger}_\mu(x) \, \hat{\pi}_D(x) \, 
i\!\stackrel{\leftrightarrow}{\partial}^{\,\mu} \! D(x) 
+ \frac{g_{_{K^\ast K \pi}}}{\sqrt{2}} \, K^{\ast \, \dagger}_\mu(x) \, 
\hat{\pi}_K(x) \, i\!\stackrel{\leftrightarrow}{\partial}^{\,\mu} \! K(x) 
\nonumber\\ 
&+& g_{_{D^\ast D_s K}} \, D^\ast_\mu(x) \, K(x)
\, i\!\stackrel{\leftrightarrow}{\partial}^{\,\mu} \! D_s^-(x)
+ g_{_{D^\ast_s D K}} 
D^{\ast \, -}_{s, \mu}(x) \, D(x)
\, i\!\stackrel{\leftrightarrow}{\partial}^{\,\mu} \! K(x) \nonumber\\
&-& i g_{_{K^\ast D_s^\ast D^\ast}} \, 
\biggl[ D^{\ast \, - \, \mu\nu }_s(x) D^{\ast}_\mu(x) K^\ast_\nu(x) 
+ D^{\ast \, \mu\nu}(x) K^\ast_\mu(x) D^{\ast \, -}_{s, \nu}(x) 
+ K^{\ast \, \mu\nu}(x) D^{\ast \, -}_{s, \mu}(x) D^\ast_\nu(x) \biggr] 
\nonumber\\
&+&g_{_{D_{s1}}} \, D_{s1}^{\mu -}(x) \, \int\! dy \, \Phi_{D_{s1}}(y^2) \, 
D^\ast_\mu(x+w_{KD^\ast} y) K(x-w_{D^\ast K} y)  \, + \, {\rm H.c.} \,, 
\en 
where summation over isospin indices is understood, 
$\Box = \partial^\mu \partial_\mu$ and 
$A \stackrel{\leftrightarrow}{\partial} B 
\equiv A \partial B - B \partial A$. Here, $\vec\pi = (\pi_1, \pi_2, \pi_3)$ 
is the triplet of pions, $\hat{\pi}_D = \pi_1 \tau_1 + \pi_2 \tau_2 
+ \pi_3 (\tau_3 \cos\varepsilon + I \sin\varepsilon/\sqrt{3})$, 
$\hat{\pi}_K = \pi_1 \tau_1 + \pi_2 \tau_2 
+ \pi_3 (\tau_3 \cos\varepsilon + I \sin\varepsilon \sqrt{3})$, 
 $D^{(\ast)}$ and $K^{(\ast)}$ are the doublets of pseudoscalar (vector) 
mesons, $D_s^\pm$ and $D_s^{\ast \, \pm}$ 
are the pseudoscalar and vector charm-strange mesons, respectively, 
$V^{\ast \, \mu\nu} = \partial^\mu V^{\ast \, \nu} - 
\partial^\nu V^{\ast \, \mu}$ is the stress tensor of vector meson field.   
In our convention the isospin-symmetric meson masses of the 
isomultiplets $m_\pi, m_P, m_V$ are identified with the masses 
of the charged partners. The quantities $\delta_{M}$ are the 
isospin-breaking parameters which are fixed by the difference of masses 
squared of the charged and neutral members of the isomultiplets as:
$\delta_M  = m_{M^\pm}^2 - m_{M^0}^2$ and
$m_{M^0} \equiv m_{\bar M^0}\,.$  
The set of mesonic masses is taken from data~\cite{Yao:2006px}.  
From Eq.~(\ref{L_strong}) it is evident that the couplings of $\pi^0$ 
to $D^\ast D$ and $K^\ast K$ mesonic pairs contain two terms -   
the ``dominant'' coupling (proportional to $\cos\varepsilon$) and 
the ``suppressed'' coupling (proportional to $\sin\varepsilon$). It means 
that the first coupling survives in the isospin limit, while the second 
one vanishes. In the context of the ``direct'' and 
``$\eta-\pi^0$ mixing'' diagrams considered by us in the preceding 
paper~\cite{Faessler:2007gv}, the first coupling generates the 
``direct'' diagrams of Fig.2 in~\cite{Faessler:2007gv}, while 
the second coupling results in the ``mixing'' diagrams of Fig.3 
in~\cite{Faessler:2007gv}. 
  
The free meson propagators are given by the standard expressions 
\eq
i \, D_M(x-y) = \langle 0 | T \, M(x) \, M^\dagger(y)  | 0 \rangle
\ = \
\int\frac{d^4k}{(2\pi)^4i} \, e^{-ik(x-y)} \ \tilde D_M(k) 
\en 
for the scalar (pseudoscalar) fields, where 
$\tilde D_M(k) = (m_M^2 - k^2 - i\epsilon)^{-1}$
and 
\eq
i \, D_{M^\ast}^{\mu\nu}(x-y) = \langle 0 | T \, M^{\ast \, \mu}(x) \,
M^{\ast \, \nu \, \dagger}(y) | 0 \rangle \ = \
\int\frac{d^4k}{(2\pi)^4i} \, e^{-ik(x-y)} \
\tilde D_{M^\ast}^{\mu\nu}(k)
\en 
for the vector (axial) fields, where
$\tilde D_{M^\ast}^{\mu\nu}(k) = ( - g^{\mu\nu} + k^\mu k^\nu/m_{M^\ast}^2) \, 
(m_{M^\ast}^2 - k^2 - i\epsilon)^{-1}\,.$

In Eq.~(\ref{L_strong}) we use the same set of strong coupling 
constants as used for the decay properties of the 
$D_{s0}^\ast$ meson (see details in Ref.~\cite{Faessler:2007gv}). 
In particular, we have 
\eq 
g_{_{D^\ast D \pi}}  = 17.9 \,, \hspace*{.15cm}  
g_{_{K^\ast K \pi}}  = 4.61 \,, \hspace*{.15cm}  
g_{_{D^\ast D_s K}}  = g_{_{K^\ast D_s D}} = 2.02 \,,  
\hspace*{.15cm} g_{_{D_s^\ast D K}} = 1.84 \,. 
\en
The additional three-vector meson coupling $g_{_{D_s^\ast D^\ast K^\ast}}$ 
is a free parameter in our calculation. It should be of order 1. 
Finally, we fix the coupling $g_{_{D_{s1}}}$,  
which is given by Eq.~(\ref{gDs1_coupling}) in terms of the adjustable 
vertex function. Using the Gaussian vertex function we obtain the result 
that this coupling is quite stable with respect to a variation of the 
scale parameter $\Lambda_{D_{s1}}$. 
In particular, varying $\Lambda_{D_{s1}}$ from 1 to 2~GeV, 
we get a range of values for $g_{_{D_{s1}}}$ from 11.62 to 10.17~GeV.  
Note, this result is in agreement with prediction of the light-cone  
QCD sum rules $g_{_{D_{s1}}} = 10.5 \pm 3.5$~GeV~\cite{Wang:2006zw}.  
In the following we also test the sensitivity of the decay properties  
of the $D_{s1}$ meson to the variation of~$\Lambda_{D_{s1}}$.  

The electromagnetic field is included in the Lagrangian (\ref{L_full}) 
using minimal substitution i.e. each derivative acting on a charged meson 
field is replaced by the covariant one:  
$\partial^\mu M^{(\ast) \pm} \, \to \, 
(\partial^\mu \mp ie A^\mu) \, M^{(\ast) \pm} \,.$
Note, that the strong $D_{s1} D^\ast K$ interaction Lagrangian should also be 
modified in order to restore electromagnetic gauge invariance. 
It proceeds in a way suggested in Ref.~\cite{Mandelstam:1962mi} and 
extensively used in Refs.~\cite{Anikin:1995cf,Faessler:2007gv}. 
In particular, each charged constituent meson field 
(i.e. $D^{\ast \pm}$ and $K^\pm$) in ${\cal L}_{D_{s1}}$ 
is multiplied by the gauge field exponential resulting in
\eq\label{L_str_gauging}
{\cal L}_{D_{s1} + {\rm em}}(x) &=& g_{_{D_{s1}}} \,
D_{s1}^{\mu \, -}(x) \, \int\! dy \, \Phi_{D_{s1}}(y^2) \,
\biggl\{ e^{- i e I(x+w_{_{KD^\ast}} y,x,P)} 
D^{\ast +}_\mu(x+w_{_{KD^\ast}} y) 
K^0(x-w_{_{D^\ast K}} y)  \, \nonumber\\
&+& D^{\ast 0}_\mu(x+w_{_{KD^\ast}} y) 
e^{- i e I(x-w_{_{D^\ast K}}y,x,P)}  
K^+(x-w_{_{D^\ast K}} y) \biggr\} \, + \, {\rm H.c.} 
\en
where
\eq\label{path}
I(x,y,P) = \int\limits_y^{x} dz_\mu A^\mu(z).
\en
For the derivative of $I(x,y,P)$ we use the
path-independent prescription suggested in~\cite{Mandelstam:1962mi}
which in turn states that the derivative of $I(x,y,P)$ does
not depend on the path $P$ originally used in the definition. 
The nonminimal substitution (\ref{L_str_gauging}) is therefore 
completely equivalent to the minimal prescription.
We should stress, that in the calculation of the amplitudes of the 
radiative $D_{s1} \to D_s \gamma$ decay,
in Eq.~(\ref{L_str_gauging}) we only need to keep terms linear in
$A_\mu$, that is the four-particle coupling
$D_{s1} D^\ast K\gamma$. Concluding the discussion of the effective 
Lagrangian we stress that all couplings occurring in the diagrams 
contributing to the decays $D_{s1} \to D_s^\ast \pi^0$ and 
$D_{s1} \to D_s \gamma$ are defined and explicitly fixed, 
except for $g_{_{D_s^\ast D^\ast K^\ast}}$.

\section{Strong and radiative decays of the $D_{s1}$ meson} 

\subsection{Matrix elements and decay widths} 

The matrix elements describing the strong $D_{s1} \to D_s^\ast \pi^0$ and 
radiative $D_{s1} \to D_s \gamma$ decays are defined as follows 
\eq\label{Minv_str} 
M( D_{s1}^+(p) \to D_s^*(p^\prime) \pi^0(q) ) =
\epsilon_\mu (p) \epsilon^{\ast}_\nu(p^\prime) \ 
( g^{\mu\nu} \ G_{D_{s1}D_s^\ast\pi}
\ - \ v^{\prime\mu} v^\nu \ F_{D_{s1}D_s^\ast\pi} )
\en 
and 
\eq\label{Minv_em}  
M( D_{s1}^+(p) \to D_s(p^\prime) \gamma(q) ) = 
\epsilon_\mu(p)\epsilon^\ast_\nu(q) 
\ (g^{\mu\nu} \ pq \ - \ q^\mu p^\nu) \ e \ G_{D_{s1}D_s\gamma}\;, 
\en 
where $v = p/m_{D_{s1}}$ and $v^\prime = p^\prime/m_{D_s^\ast}$ 
are the four-velocities of the $D_{s1}$ and $D_s^\ast$ mesons, 
$G(F)_{D_{s1}D_s^\ast\pi}$ and $G_{D_{s1}D_s\gamma}$ are the 
corresponding effective coupling constants.
The coherent sum of all the diagrams in Fig.3 contributing 
to the radiative decay $D_{s1} \to D_s \gamma$ is gauge
invariant, while the contribution of each diagram is definitely not gauge
invariant. As in Ref.~\cite{Faessler:2007gv}, for convenience we split 
each individual diagram into a gauge-invariant piece and a remainder 
proportional to the Lorentz structure $g_{\mu\nu}$, which is noninvariant.
One can prove that the sum of the noninvariant terms vanishes
due to gauge invariance. In the following discussion of the numerical
results we will only deal with the gauge-invariant contribution of 
the separate diagrams of Fig.3.

Using Eqs.~(\ref{Minv_str}) and (\ref{Minv_em}) the strong 
$D_{s1} \to D_s^\ast \pi^0$ and radiative $D_{s1} \to D_s \gamma$ 
decay widths are calculated according to the expressions: 
\eq
\Gamma(D_{s1} \to D_s^\ast \pi) &=&
\frac{P^\ast_{\pi^0}}{12 \pi m_{D_{s1}}^2} \, 
\biggl\{ G_{D_{s1}D_s^\ast\pi}^2 + \frac{1}{2} 
\biggl( G_{D_{s1}D_s^\ast\pi} \, w - F_{D_{s1}D_s^\ast\pi} 
(w^2 - 1) \biggr)^2 \biggr\} \,, \nonumber\\ 
\Gamma(D_{s1} \to D_s \gamma) &=& \frac{\alpha}{3} \,
G_{D_{s1} D_s \gamma}^2 \, P_\gamma^{\ast \, 3}
\en
where $w = v v^\prime = (m_{D_{s1}}^2 + m_{D_{s^\ast}}^2 
- m_\pi^2)/(2 m_{D_{s1}} m_{D_s^\ast})$; $P^\ast_{\pi^0} = 
\lambda^{1/2}(m_{D_{s1}}^2,m_{D_s^\ast}^2,m_{\pi^0}^2)/(2 m_{D_{s1}})$ 
and $P^\ast_\gamma = (m_{D_{s1}}^2 - m_{D_{s}}^2)/(2 m_{D_{s1}})$ 
are the corresponding three-momenta of the decay products with 
$\lambda(x,y,z) = x^2 + y^2 + z^2 - 2xy - 2xz - 2yz$ being 
the K\"allen function.

\subsection{Numerical results} 

First we discuss the results for the strong decay $D_{s1} \to D_s^\ast \pi^0$.
Here the main contribution to the decay width comes from the diagrams
of Figs.2(c) and 2(d), while the contribution of the diagrams
of Figs.2(a) and 2(b) is relatively suppressed by a factor of $\sim 10^{-2}$. 
On the other hand, the contribution to the decay width generated 
by the dominant coupling (proportional to $\cos\varepsilon$) exceeds 
the contribution due to the suppressed coupling (proportional to 
$\sin\varepsilon$) by a factor of 2 on average  
[see Lagrangian (12) and discussion in Sec.II]. 
The ``dominant'' coupling corresponds to the 
``direct'' diagrams of Fig.2 in~\cite{Faessler:2007gv}, while 
the second coupling concerns the ``mixing'' diagrams of Fig.3 
in~\cite{Faessler:2007gv}. In this paper, we combine these two types 
of diagrams together using the unitary transformation~(\ref{Unit_trans}).    

In terms of the unknown dimensionless coupling constant 
$g_{_{D_s^\ast D^\ast K^\ast}}$ the results 
for $\Gamma(D_{s1} \to D_s^\ast \pi)$ range from 
$(14.8 \, g_{_{D_s^\ast D^\ast K^\ast}}^2$) keV at $\Lambda_{D_{s1}} = 1$ GeV 
to $(23.4 \, g_{_{D_s^\ast D^\ast K^\ast}}^2$) keV 
at $\Lambda_{D_{s1}} = 2$ GeV. 
Choosing a typical value of
$g_{_{D_s^\ast D^\ast K^\ast}} \simeq g_{_{D_s^\ast D K}} = 1.84$, we get
\eq
\Gamma(D_{s1} \to D_s^\ast \pi) = 50.1 - 79.2 \ {\rm keV}
\en
where the range of values for our results is due to the 
variation of $\Lambda_{D_{s1}}$ from 1 to 2 GeV. 
An increase of $\Lambda_{D_{s1}}$ leads to an increase of the width.
In Table 1 we present our results for the decay width
$\Gamma(D_{s1} \to D_s^\ast \pi)$ including a variation of the scale
parameter $\Lambda_{D_{s1}}$ from 1 to 2 GeV and compare
them to previous theoretical predictions. 

Now we turn to the discussion of the radiative decay
$D_{s1} \to D_s \gamma$. By construction, using
a gauge-invariant and Lorentz-covariant effective Lagrangian, the full
amplitude for this process is gauge-invariant, while the separate
contributions of the different diagrams of Fig.3 are not.
It is important to stress that the diagrams of Fig.3  fall into
two separately gauge-invariant sets: one set includes the diagrams
of Figs.3(a), 3(c), 3(e), and 3(g) (with loops containing virtual 
$D^{\ast +}$ and $K^0$ mesons), generated by the coupling of 
$D_{s1}$ to the $D^{\ast +}$ and $K^0$ constituents. 
Another set contains the diagrams of Figs.3(b), 3(d), 3(f), and 3(h) 
(with loops containing virtual $D^{\ast 0}$ and $K^+$ mesons) with 
the coupling of $D_{s1}$ to $D^{\ast 0}$ and $K^+$. 

First we give the results for the effective coupling constant
$G_{D_{s1} D_s \gamma}$: the total result and partial
contributions of the different diagrams of Fig.3
(marked by 3(a), 3(b), etc.). In the analysis of the electromagnetic
decay $D_{s1} \to D_s \gamma$ we restrict to the isospin limit,
i.e. we do not include the isospin-breaking effects in the meson masses and
proceed with the masses of the charged particles. In the isospin limit
the diagrams of Fig.3(g) and 3(h) are equal to each other.
For a value of $\Lambda_{D_{s1}} = 1$ GeV we get 
(here we only deal with the gauge-invariant parts of the diagrams of 
Figs.3(a)-3(d), 3(g), and 3(h), while the gauge-invariant parts of 
diagrams in Figs.3(e) and 3(f) are equal to zero):   
\eq
& &G_{D_{s1} D_s \gamma} = 0.106 \ {\rm GeV}^{-1} \,,
\hspace*{.3cm}
G_{D_{s1} D_s \gamma}^{3a} = 0.008 \ {\rm GeV}^{-1} \,,
\hspace*{.3cm}
G_{D_{s1} D_s \gamma}^{3b} = 0.090 \ {\rm GeV}^{-1} \,, \\
& &G_{D_{s1} D_s \gamma}^{3c} = 7 \times 10^{-4} \ {\rm GeV}^{-1} \,,
\hspace*{.4cm}
G_{D_{s1} D_s \gamma}^{3d} = - 2 \times 10^{-4} \ {\rm GeV}^{-1} \,,
\hspace*{.7cm}
G_{D_{s1} D_s \gamma}^{3g} \equiv G_{D_{s1} D_s \gamma}^{3h} =
0.004 \ {\rm GeV}^{-1} \,. \nonumber
\en
From the results it is clear that the contact diagrams of Fig.3(c)  
and 3(d) are strongly suppressed, however these diagrams are kept to guarantee 
gauge invariance. The main contribution comes from the diagram of 
Fig.3(b) where the photon couples to the $K^+$. Note, that 
the same conclusion concerning the dominance of the triangle diagram with the 
$\gamma K^+$ coupling was obtained in the analysis of  
the radiative decay of the $D_{s0}^\ast(2317)$ into 
$D_s \gamma$~\cite{Faessler:2007gv}. 

The result for the decay width of the transition
$D_{s1} \to D_s \gamma$ is:
\eq
\Gamma(D_{s1} \to D_s \gamma) = 2.37 \ {\rm keV}\,. 
\en 

In Table 2 we summarize our results for 
$\Gamma(D_{s1} \to D_s \gamma)$ including a variation of the scale
parameter $\Lambda_{D_{s1}}$ from 1 to 2 GeV (an increase of
$\Lambda_{D_{s1}}$ leads to a larger value for the width). 
We also compare to predictions of other theoretical approaches.
Our results have a minor dependence on the parameter
$\Lambda_{D_{s1}}$ and are also in agreement
with previous calculations. 
Finally, in Table 3 we present a comparison of our results 
for the ratio $R = \Gamma(D_{s1} \to D_s \gamma/\Gamma(D_{s1} 
\to D_s^\ast \pi)$ of the radiative and strong decay of $D_{s1}$ 
mesons with data and other approaches. Our result for $R \simeq 0.05$ 
is insensitive to a variation of the size parameter $\Lambda_{D_{s1}}$. 

\section{Summary}

We studied the new charm-strange meson $D_{s1}(2460)$
in the hadronic molecule interpretation, considering a bound state
of $D^\ast$ and $K$ mesons. Using an effective Lagrangian approach
we calculated the strong $D_{s1} \to D_s^\ast \pi^0$ and
radiative $D_{s1} \to D_s \gamma$ decays. 
An important consequence of the $D^\ast K$ molecular structure
of the $D_{s1}(2460)$ meson is that the presence of $u(d)$ quarks in
the $D^\ast$ and $K$ meson loops gives rise to a direct strong 
isospin-violating transition $D_{s1} \to D_s^\ast \pi^0$ in addition to
the decay mechanism induced by $\eta-\pi^0$ mixing as was considered
previously in literature. We found that the direct transition 
dominates over the $\eta-\pi^0$ mixing transition. In the present paper 
we combined these two mechanisms to the modified pion-emission diagrams
due to the diagonalization of the pseudoscalar meson mass 
term~\cite{Gasser:1984gg} (see also Eq.~(\ref{Unit_trans})).   
In the language of the strong effective Lagrangian~(\ref{L_strong}) 
the ``direct'' mechanism corresponds to the $K K^\ast \pi^0$ and 
$D D^\ast \pi^0$ couplings containing the $\tau_3 \cos\varepsilon$ 
flavor matrix, while the ``mixing'' mechanism corresponds to the  
$K K^\ast \pi^0$ and $D D^\ast \pi^0$ couplings containing the 
$I \cos\varepsilon$ flavor matrix. Finally, the main contribution to 
the strong decay $D_{s1} \to D_s^\ast \pi^0$ width  
arises from the diagrams of Figs.2(c) and 2(d).  
The contribution of the diagrams of Figs.2(a) and 2(b) to the decay 
width is suppressed  by a factor of $\sim 10^{-2}$. 
In the case of the radiative decay $D_{s1} \to D_s \gamma$ the dominant 
contribution comes from the diagram of Fig.3(b). 

Our results for the partial decay widths of the $D_{s1}(2460)$ and their ratio 
are summarized as follows:
\eq
& &\Gamma(D_{s1} \to D_s^\ast \pi) = 50.1 - 79.2 \ {\rm keV}\,, \nonumber\\
& &\Gamma(D_{s1} \to D_s \gamma) = 2.37 - 3.73 \ {\rm keV}\,,\nonumber\\
& &R = \frac{\Gamma(D_{s1} \to D_s \gamma}{\Gamma(D_{s1} \to D_s^\ast \pi)}  
\simeq 0.05 \,. 
\en

\begin{acknowledgments}

This work was supported by the DFG under contracts FA67/31-1 and
GRK683. This research is also part of the EU Integrated
Infrastructure Initiative Hadronphysics project under contract
number RII3-CT-2004-506078 and President grant of Russia
"Scientific Schools"  No. 5103.2006.2.

\end{acknowledgments}

\newpage 

\begin{table}
\begin{center}
{\bf Table 1.} 
Decay width of $D_{s1} \to D_s^\ast \pi^0$. \\ 
For our result the range of values is due \\ to the 
variation of $\Lambda_{D_{s1}}$ from 1 to 2 GeV.

\vspace*{.25cm}

\def\arraystretch{1.2}
\begin{tabular}{|l|l|}
\hline
\hspace*{.5cm}
Approach \hspace*{.5cm}
& \hspace*{.5cm}
$\Gamma(D_{s1} \to D_s^\ast \pi^0)$ (keV) \hspace*{.5cm} \\
\hline
\,\,\,\,\,
Ref.~\cite{Colangelo:2003vg}
\,\,\,\,\, & \,\,\,\,\,\,\,\, 7 $\pm$ 1 \,\,\,\,\, \\
\,\,\,\,\,
Ref.~\cite{Godfrey:2003kg}
\,\,\,\,\, & \,\,\,\,\,\,\,\, 10 \,\,\,\,\, \\
\,\,\,\,\,
Ref.~\cite{Guo:2006rp}
& \,\,\,\,\,\,\,\, 11.41 \,\,\,\,\, \\
\,\,\,\,\,
Ref.~\cite{Bardeen:2003kt}
\,\,\,\,\, & \,\,\,\,\,\,\,\, 21.5 \,\,\,\,\, \\
\,\,\,\,\,
Ref.~\cite{Fayyazuddin:2003dp}
& \,\,\,\,\,\,\,\, 32 \,\,\,\,\, \\
\,\,\,\,\,
Ref.~\cite{Lu:2006ry}
& \,\,\,\,\,\,\,\, 35 \,\,\,\,\, \\
\,\,\,\,\,
Ref.~\cite{Wei:2005ag}
\,\,\,\,\,       & \,\,\,\,\,\,\,\, 43 $\pm$ 8 \,\,\,\,\, \\
\,\,\,\,\,
Ref.~\cite{Lutz:2007er}
\,\,\,\,\, & \,\,\,\,\,\,\,\, 140   \,\,\,\,\, \\
\,\,\,\,\,
Ref.~\cite{Ishida:2003gu}
\,\,\,\,\, & \,\,\,\,\,\,\,\, 155 $\pm$ 70 \,\,\,\,\, \\
\,\,\,\,\,
Ref.~\cite{Azimov:2004xk}
\,\,\,\,\, & \,\,\,\,\,\,\,\, 187 $\pm$ 73 \,\,\,\,\, \\
\,\,\,\,\, Our result 
\,\,\,\,\, & \,\,\,\,\,\,\,\, 50.1 $-$ 79.2 \,\,\,\,\, \\
\hline
\end{tabular}
\end{center}
\end{table}

\begin{table}
\begin{center}
{\bf Table 2.}
Decay width of $D_{s1} \to D_s \gamma \ $. \\ 
For our result the range of values is due \\ to the 
variation of $\Lambda_{D_{s1}}$ from 1 to 2 GeV.

\vspace*{1cm}

\def\arraystretch{1.2}
\begin{tabular}{|l|l|}
\hline
\hspace*{.5cm}
Approach \hspace*{.5cm}
& \hspace*{.5cm}
$\Gamma(D_{s1} \to D_s \gamma) \ \ $ (keV) \hspace*{.5cm} \\
\hline
\,\,\,\,\,
Ref.~\cite{Liu:2006jx}
\,\,\,\,\, & \,\,\,\,\,\,\,\, 0.6 - 2.9 \,\,\,\,\, \\
\,\,\,\,\,
Ref.~\cite{Azimov:2004xk}
\,\,\,\,\, & \,\,\,\,\,\,\,\, $\approx$ 2 \,\,\,\,\, \\ 
\,\,\,\,\,
Ref.~\cite{Bardeen:2003kt}
\,\,\,\,\, & \,\,\,\,\,\,\,\, 5.08 \,\,\,\,\, \\
\,\,\,\,\,
Ref.~\cite{Wang:2006mf}
\,\,\,\,\, & \,\,\,\,\,\,\,\, 5.5 $-$ 31.2 \,\,\,\,\, \\
\,\,\,\,\,
Ref.~\cite{Godfrey:2003kg}
\,\,\,\,\, & \,\,\,\,\,\,\,\, 6.2 \,\,\,\,\, \\
\,\,\,\,\,
Ref.~\cite{Close:2005se}
\,\,\,\,\, & \,\,\,\,\,\,\,\, $\le$ 7.3 \,\,\,\,\, \\
\,\,\,\,\,
Ref.~\cite{Colangelo:2005hv}
& \,\,\,\,\,\,\,\, 19 $-$ 29 \,\,\,\,\, \\
\,\,\,\,\,
Ref.~\cite{Lutz:2007er}
\,\,\,\,\, & \,\,\,\,\,\,\,\, $\simeq$ 43.6   \,\,\,\,\, \\
\,\,\,\,\,
Ref.~\cite{Ishida:2003gu}
\,\,\,\,\, & \,\,\,\,\,\,\,\, 93 \,\,\,\,\, \\
\,\,\,\,\, Our result 
& \,\,\,\,\,\,\,\, 2.37 $-$ 3.73 \,\,\,\,\, \\
\hline
\end{tabular}
\end{center}
\end{table}

\begin{table}
\begin{center}
{\bf Table 3.} 
The ratio  
$R = \Gamma(D_{s1} \to D_s \gamma)/\Gamma(D_{s1} \to D_s^\ast \pi)$. \\ 
For our result the range of values is due \\ to the 
variation of $\Lambda_{D_{s1}}$ from 1 to 2 GeV. 

\vspace*{.25cm}

\def\arraystretch{1.2}
\begin{tabular}{|l|l|}
\hline
\hspace*{.5cm}
Approach \hspace*{.5cm}
& \hspace*{.5cm}
$R$ \hspace*{.5cm} \\
\hline 
\,\,\,\,\,
Ref.~\cite{Azimov:2004xk}
\,\,\,\,\, & \,\,\,\,\,\,\,\, 0.01 - 0.02 \,\,\,\,\, \\
\,\,\,\,\,
Ref.~\cite{Bardeen:2003kt}
\,\,\,\,\, & \,\,\,\,\,\,\,\, 0.24  \,\,\,\,\, \\
\,\,\,\,\,
Ref.~\cite{Lutz:2007er}
\,\,\,\,\, & \,\,\,\,\,\,\,\, $\simeq$ 0.31   \,\,\,\,\, \\
\,\,\,\,\,
Ref.~\cite{Ishida:2003gu}
\,\,\,\,\, & \,\,\,\,\,\,\,\, 0.41 - 1.09  \,\,\,\,\, \\
\,\,\,\,\,
Ref.~\cite{Godfrey:2003kg}
\,\,\,\,\, & \,\,\,\,\,\,\,\, 0.62 \,\,\,\,\, \\
\,\,\,\,\,
Data~\cite{Yao:2006px}  
\,\,\,\,\, & \,\,\,\,\,\,\,\, 0.44 $\pm$ 0.09   \,\,\,\,\, \\
\,\,\,\,\, Our result  
& \,\,\,\,\,\,\,\, $\simeq$ 0.05 \,\,\,\,\, \\
\hline
\end{tabular}
\end{center}
\end{table}

\newpage 

\begin{figure}

\vspace*{2cm} 

\begin{center}
\epsfig{file=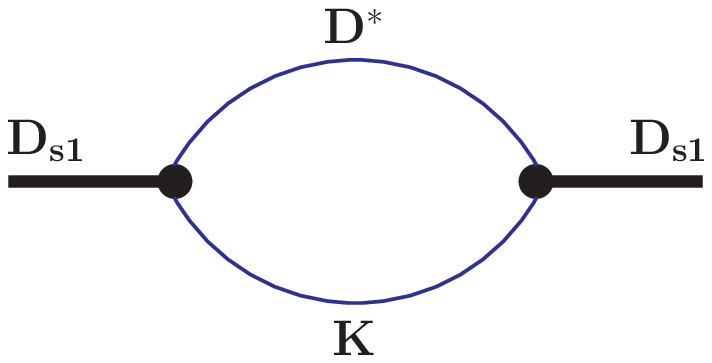, scale=.5}
\end{center}
\caption{Mass operator of the $D_{s1}(2460)$ meson.}

\vspace*{2cm}

\begin{center}
\epsfig{file=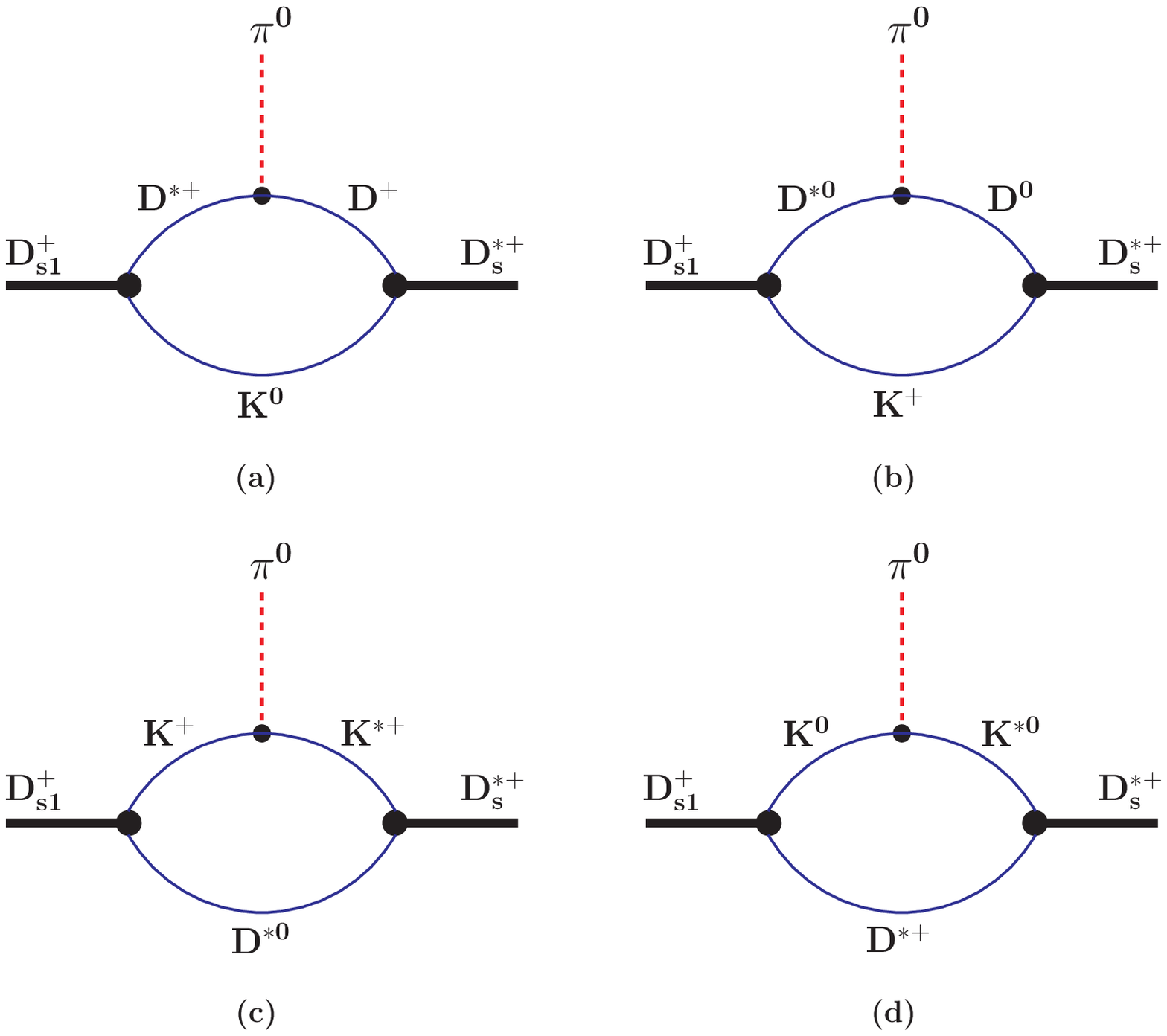, scale=.5}
\end{center}
\caption{Diagrams contributing to the strong transition
$D_{s1}^{+} \to D_{s}^{\ast +} + \pi^0$.}
\end{figure}

\newpage

\begin{figure}

\vspace*{2cm}

\begin{center}
\epsfig{file=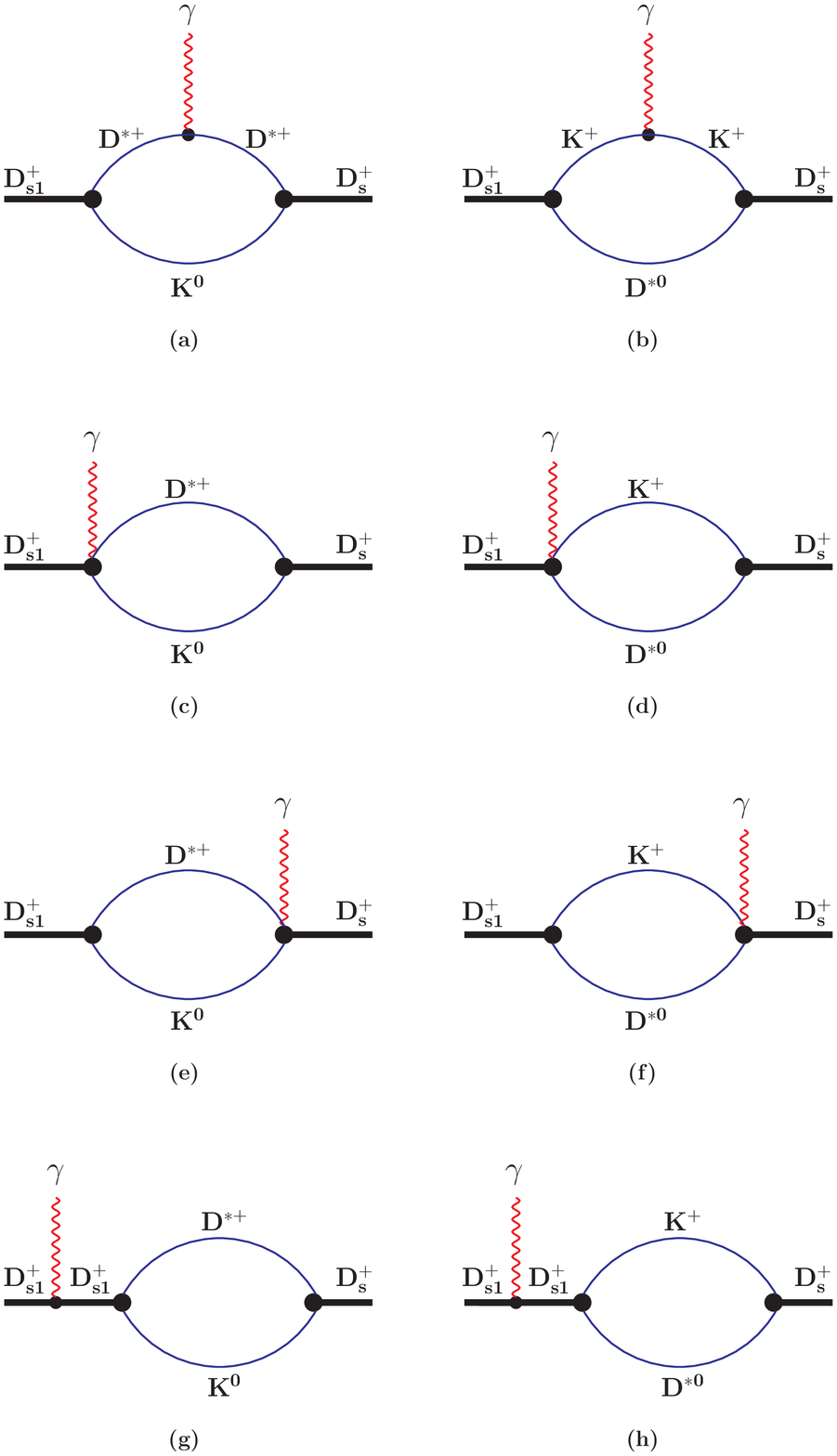, scale=.5}
\end{center}
\caption{Diagrams contributing to the radiative transition
$D_{s1}^{+} \to D_{s}^{+} + \gamma$.}
\end{figure}


\newpage

\begin{thebibliography}{99}

%\cite{Rosner:2006vc}
\bibitem{Rosner:2006vc}
  J.~L.~Rosner,
  %``Effects of S-wave thresholds,''
  Phys.\ Rev.\ D {\bf 74}, 076006 (2006)
  [arXiv:hep-ph/0608102].
  %%CITATION = HEP-PH 0608102;%%
%\cite{Barnes:2003dj}
\bibitem{Barnes:2003dj}
  T.~Barnes, F.~E.~Close and H.~J.~Lipkin,
  %``Implications of a D K molecule at 2.32-GeV,''
  Phys.\ Rev.\ D {\bf 68}, 054006 (2003)
  [arXiv:hep-ph/0305025].
  %%CITATION = HEP-PH 0305025;%
%\cite{Aubert:2003fg}
\bibitem{Aubert:2003fg}
  B.~Aubert {\it et al.}  [BABAR Collaboration],
  %``Observation of a narrow meson decaying to
  %$D_s^+ \pi^0$ at a mass of 2.32-GeV/c$^2$,''
  Phys.\ Rev.\ Lett.\  {\bf 90}, 242001 (2003)
  [arXiv:hep-ex/0304021].
  %%CITATION = HEP-EX 0304021;%%
%\cite{Besson:2003cp}
\bibitem{Besson:2003cp}
  D.~Besson {\it et al.}  [CLEO Collaboration],
  %``Observation of a narrow resonance of mass
  %2.46-GeV/c**2 decaying to  D/s*+
  %pi0 and confirmation of the D/sJ*(2317) state,''
  Phys.\ Rev.\ D {\bf 68}, 032002 (2003)
  [arXiv:hep-ex/0305100].
  %%CITATION = HEP-EX 0305100;%%
%\cite{Abe:2003jk}
\bibitem{Abe:2003jk}
  K.~Abe {\it et al.},
  %``Measurements of the D/sJ resonance properties,''
  Phys.\ Rev.\ Lett.\  {\bf 92} (2004) 012002
  [arXiv:hep-ex/0307052].
  %%CITATION = HEP-EX 0307052;%%
%\cite{Yao:2006px}
\bibitem{Yao:2006px}
  W.~M.~Yao {\it et al.}  [Particle Data Group],
  %``Review of particle physics,''
  J.\ Phys.\ G {\bf 33}, 1 (2006) 
  and 2007 partial update for the 2008 edition. 
  %%CITATION = JPHGB,G33,1;%%
%\cite{Colangelo:2004vu}
\bibitem{Colangelo:2004vu}
  P.~Colangelo, F.~De Fazio and R.~Ferrandes,
  %``Excited charmed mesons: Observations, analyses and puzzles,''
  Mod.\ Phys.\ Lett.\  A {\bf 19}, 2083 (2004)
  [arXiv:hep-ph/0407137].
  %%CITATION = MPLAE,A19,2083;%%
%\cite{Cheng:2003kg}
\bibitem{Cheng:2003kg}
  H.~Y.~Cheng and W.~S.~Hou,
  %``B decays as spectroscope for charmed four-quark states,''
  Phys.\ Lett.\ B {\bf 566}, 193 (2003)
  [arXiv:hep-ph/0305038].
  %%CITATION = HEP-PH 0305038;%%
%\cite{Bardeen:2003kt}
\bibitem{Bardeen:2003kt}
  W.~A.~Bardeen, E.~J.~Eichten and C.~T.~Hill,
  %``Chiral multiplets of heavy-light mesons,''
  Phys.\ Rev.\ D {\bf 68}, 054024 (2003)
  [arXiv:hep-ph/0305049].
  %%CITATION = HEP-PH 0305049;%%
%\cite{Godfrey:2003kg}
\bibitem{Godfrey:2003kg}
  S.~Godfrey,
  %``Using radiative transitions to test the 1(3)P(0)(c anti-s)
  %nature of the D/sJ*(2317)+ state,''
  Phys.\ Lett.\ B {\bf 568}, 254 (2003)
  [arXiv:hep-ph/0305122].
  %%CITATION = HEP-PH 0305122;%%
%\cite{Colangelo:2003vg}
\bibitem{Colangelo:2003vg}
  P.~Colangelo and F.~De Fazio,
  %``Understanding D/sJ(2317),''
  Phys.\ Lett.\ B {\bf 570}, 180 (2003)
  [arXiv:hep-ph/0305140].
  %%CITATION = HEP-PH 0305140;%%
%\cite{Fayyazuddin:2003dp}
\bibitem{Fayyazuddin:2003dp}
  Fayyazuddin and Riazuddin,
  %``Some comments on narrow resonances D/s1* (2.46-GeV/c**2) and D/s0
  %(2.317-GeV/c**2),''
  Phys.\ Rev.\  D {\bf 69}, 114008 (2004)
  [arXiv:hep-ph/0309283].
  %%CITATION = PHRVA,D69,114008;%%
%\cite{Ishida:2003gu}
\bibitem{Ishida:2003gu}
  S.~Ishida, M.~Ishida, T.~Komada, T.~Maeda, M.~Oda,
  K.~Yamada and I.~Yamauchi,
  %``The D/s(2317) and D/s(2463) mesons as scalar and
  %axial-vector chiralons  in
  %the covariant level-classification scheme,''
  AIP Conf.\ Proc.\  {\bf 717}, 716 (2004)
  [arXiv:hep-ph/0310061].
  %%CITATION = HEP-PH 0310061;%%
%\cite{Azimov:2004xk}
\bibitem{Azimov:2004xk}
  Y.~I.~Azimov and K.~Goeke,
  %``Decay properties of new D-mesons,''
  Eur.\ Phys.\ J.\ A {\bf 21}, 501 (2004)
  [arXiv:hep-ph/0403082].
  %%CITATION = HEP-PH 0403082;%
%\cite{Colangelo:2005hv}
\bibitem{Colangelo:2005hv}
  P.~Colangelo, F.~De Fazio and A.~Ozpineci,
  %``Radiative transitions of D/sJ*(2317) and D/sJ(2460),''
  Phys.\ Rev.\  D {\bf 72}, 074004 (2005)
  [arXiv:hep-ph/0505195].
  %%CITATION = PHRVA,D72,074004;%%
%\cite{Close:2005se}
\bibitem{Close:2005se}
  F.~E.~Close and E.~S.~Swanson,
  %``Dynamics and decay of heavy-light hadrons,''
  Phys.\ Rev.\ D {\bf 72}, 094004 (2005)
  [arXiv:hep-ph/0505206].
  %%CITATION = HEP-PH 0505206;%%
%\cite{Wei:2005ag}
\bibitem{Wei:2005ag}
  W.~Wei, P.~Z.~Huang and S.~L.~Zhu,
  %``Strong decays of D/sJ(2317) and D/sJ(2460),''
  Phys.\ Rev.\ D {\bf 73}, 034004 (2006)
  [arXiv:hep-ph/0510039].
  %%CITATION = HEP-PH 0510039;%%
%\cite{Hayashigaki:2004st}
\bibitem{Hayashigaki:2004st}
  A.~Hayashigaki and K.~Terasaki,
  %``Isospin quantum number of D/s0(2317)+,''
  Prog.\ Theor.\ Phys.\  {\bf 114}, 1191 (2005)
  [arXiv:hep-ph/0410393]. 
  %%CITATION = PTPKA,114,1191;%%
%\cite{Liu:2006jx}
\bibitem{Liu:2006jx}
  X.~Liu, Y.~M.~Yu, S.~M.~Zhao and X.~Q.~Li,
  %``Study on decays of D/sJ*(2317) and D/sJ(2460) 
  %in terms of the CQM model,''
  Eur.\ Phys.\ J.\  C {\bf 47}, 445 (2006)
  [arXiv:hep-ph/0601017].
  %%CITATION = EPHJA,C47,445;%%
%\cite{Lu:2006ry}
\bibitem{Lu:2006ry}
  J.~Lu, X.~L.~Chen, W.~Z.~Deng and S.~L.~Zhu,
  %``Pionic decays of D/sj(2317), D/sj(2460) and B/sj(5718), B/sj(5765),''
  Phys.\ Rev.\ D {\bf 73}, 054012 (2006)
  [arXiv:hep-ph/0602167].
  %%CITATION = HEP-PH 0602167;%%
%\cite{Guo:2006fu}
\bibitem{Guo:2006fu}
  F.~K.~Guo, P.~N.~Shen, H.~C.~Chiang and R.~G.~Ping,
  %``Dynamically generated 0+ heavy mesons in a heavy chiral unitary
  %approach,''
  Phys.\ Lett.\  B {\bf 641}, 278 (2006)
  [arXiv:hep-ph/0603072].
  %%CITATION = PHLTA,B641,278;%%
%\cite{Guo:2006rp}
\bibitem{Guo:2006rp}
  F.~K.~Guo, P.~N.~Shen and H.~C.~Chiang,
  %``Dynamically generated 1+ heavy mesons,''
  Phys.\ Lett.\  B {\bf 647}, 133 (2007)
  [arXiv:hep-ph/0610008].
  %%CITATION = PHLTA,B647,133;%%
%\cite{Wang:2006mf} 
\bibitem{Wang:2006mf}
  Z.~G.~Wang,
  %``Radiative decays of the $D_{s0}(2317)$, $D_{s1}(2460)$ and the related
  %strong coupling constants,''
  Phys.\ Rev.\  D {\bf 75}, 034013 (2007)
  [arXiv:hep-ph/0612225].
  %%CITATION = PHRVA,D75,034013;%%
%\cite{Faessler:2007gv}
\bibitem{Faessler:2007gv}
  A.~Faessler, T.~Gutsche, V.~E.~Lyubovitskij and Y.~L.~Ma,
  %``Strong and radiative decays of the Ds0*(2317) meson in the DK-molecule
  %picture,''
  Phys.\ Rev.\  D {\bf 76}, 014005 (2007)
  [arXiv:0705.0254 [hep-ph]].
  %%CITATION = PHRVA,D76,014005;%% 
%\cite{Gamermann:2007bm}
\bibitem{Gamermann:2007bm}
  D.~Gamermann, L.~R.~Dai and E.~Oset,
  %``Radiative decay of the dynamically generated open and hidden charm scalar
  %meson resonances D_{s0}^*(2317) and X(3700),''
  Phys.\ Rev.\  C {\bf 76}, 055205 (2007)
  [arXiv:0709.2339 [hep-ph]].
  %%CITATION = PHRVA,C76,055205;%%
%\cite{Gamermann:2007er}
\bibitem{Gamermann:2007er} 
  D.~Gamermann, 
  Talk given at the 29th Course ``Quarks in Hadrons and Nuclei'', 
  September 16--24, 2007, Erice, Sicily. 
%\cite{Lutz:2007er}
\bibitem{Lutz:2007er}
  M.~F.~M.~Lutz, 
  Talks given at the 11th International Conference on Meson-Nucleon Physics 
  and the Structure of the Nucleon (MENU2007), September 10--14, 2007,
  J\"ulich, Germany and International School of Nuclear Physics, 
  29th Course ``Quarks in Hadrons and Nuclei'', 
  September 16--24, 2007, Erice, Sicily. 
%\cite{Lyubovitskij:2007er}
\bibitem{Lyubovitskij:2007er} 
  V.~E.~Lyubovitskij, 
  Talks given at the 11th International Conference on Meson-Nucleon Physics 
  and the Structure of the Nucleon (MENU2007), September 10--14, 2007,
  J\"ulich, Germany and International School of Nuclear Physics, 
  29th Course ``Quarks in Hadrons and Nuclei'', 
  September 16--24, 2007, Erice, Sicily. 
%\cite{Faessler:2007cu}
\bibitem{Faessler:2007cu}
  A.~Faessler, T.~Gutsche, S.~Kovalenko and V.~E.~Lyubovitskij,
  %``D/s0*(2317) and D/s1(2460) mesons in two-body B-meson decays,''
  Phys.\ Rev.\  D {\bf 76}, 014003 (2007)
  [arXiv:0705.0892 [hep-ph]].
  %%CITATION = PHRVA,D76,014003;%%
%\cite{Weinberg:1962hj}\ 
\bibitem{Weinberg:1962hj}
  S.~Weinberg,
  %``Elementary Particle Theory Of Composite Particles,''
  Phys.\ Rev.\  {\bf 130}, 776 (1963);
  %%CITATION = PHRVA,130,776;%%
%\cite{Salam:1962ap}
%\bibitem{Salam:1962ap}
  A.~Salam,
  %``Lagrangian Theory Of Composite Particles,''
  Nuovo Cim.\  {\bf 25}, 224 (1962);
  %%CITATION = NUCIA,25,224;%%
%\cite{Hayashi:1967hk}
  %\bibitem{Hayashi:1967hk}
  K.~Hayashi, M.~Hirayama, T.~Muta, N.~Seto and T.~Shirafuji,
  Fortsch.\ Phys.\ {\bf 15}, 625 (1967).
  %%CITATION = FPYKA,15,625;%%
%\cite{Efimov:1993ei} 
\bibitem{Efimov:1993ei}
  G.~V.~Efimov and M.~A.~Ivanov,
  {\it The Quark Confinement Model of Hadrons},
  (IOP Publishing, Bristol $\&$ Philadelphia, 1993). 
%\cite{Anikin:1995cf}
\bibitem{Anikin:1995cf}
  I.~V.~Anikin, M.~A.~Ivanov, N.~B.~Kulimanova and V.~E.~Lyubovitskij,
  %``The Extended Nambu-Jona-Lasinio model with separable interaction:
  %Low-energy pion physics and pion nucleon form-factor,''
  Z.\ Phys.\ C {\bf 65}, 681 (1995);
  %%CITATION = ZEPYA,C65,681;%%
%\cite{Ivanov:1996pz}
%\bibitem{Ivanov:1996pz}
  M.~A.~Ivanov, M.~P.~Locher and V.~E.~Lyubovitskij,
  %``Electromagnetic form factors of nucleons in a relativistic
  %three-quark model,''
  Few Body Syst.\  {\bf 21}, 131 (1996); 
%\cite{Ivanov:1996fj}
%\bibitem{Ivanov:1996fj}
  M.~A.~Ivanov, V.~E.~Lyubovitskij, J.~G.~K\"orner and P.~Kroll,
  %``Heavy baryon transitions in
  %a relativistic three-quark model,''
  Phys.\ Rev.\ D {\bf 56}, 348 (1997)
  [arXiv:hep-ph/9612463];
  %%CITATION = HEP-PH 9612463;%%
%\cite{Ivanov:1999bk}
%\bibitem{Ivanov:1999bk}
  M.~A.~Ivanov, J.~G.~K\"orner, V.~E.~Lyubovitskij and A.~G.~Rusetsky, 
  %``Strong and radiative decays of heavy flavored baryons,'' 
  Phys.\ Rev.\ D {\bf 60}, 094002 (1999) 
  [arXiv:hep-ph/9904421]; 
  %%CITATION = HEP-PH 9904421;%%
%\cite{Faessler:2003yf}
%\bibitem{Faessler:2003yf}
  A.~Faessler, T.~Gutsche, M.~A.~Ivanov, V.~E.~Lyubovitskij and P.~Wang, 
  %``Pion and sigma meson properties in a relativistic quark model,'' 
  Phys.\ Rev.\  D {\bf 68}, 014011 (2003) 
  [arXiv:hep-ph/0304031];  
  %%CITATION = PHRVA,D68,014011;%% 
%\cite{Faessler:2006ft}
%\bibitem{Faessler:2006ft}
  A.~Faessler, T.~Gutsche, M.~A.~Ivanov, J.~G.~Korner,
  V.~E.~Lyubovitskij, D.~Nicmorus and K.~Pumsa-ard,
  %``Magnetic moments of heavy baryons in the relativistic
  %three-quark model,''
  Phys.\ Rev.\ D {\bf 73}, 094013 (2006)
  [arXiv:hep-ph/0602193]; 
  %%CITATION = HEP-PH 0602193;%%
%\cite{Faessler:2006ky}
%\bibitem{Faessler:2006ky}
  A.~Faessler, T.~Gutsche, B.~R.~Holstein, V.~E.~Lyubovitskij,
  D.~Nicmorus and K.~Pumsa-ard,
  %``Light baryon magnetic moments and N --> Delta gamma
  %transition in a Lorentz covariant chiral quark approach,''
  Phys.\ Rev.\ D {\bf 74}, 074010 (2006)
  [arXiv:hep-ph/0608015].
  %%CITATION = HEP-PH 0608015;%%
%\cite{Gasser:1984gg}
\bibitem{Gasser:1984gg}
  J.~Gasser and H.~Leutwyler,
  %``Chiral Perturbation Theory: Expansions 
  %In The Mass Of The Strange Quark,''
  Nucl.\ Phys.\  B {\bf 250}, 465 (1985).
  %%CITATION = NUPHA,B250,465;%%
%\cite{Cho:1994zu}
\bibitem{Cho:1994zu}
  P.~L.~Cho and M.~B.~Wise,
  %``Comment On D(S)* $\to$ D(S) Pi0 Decay,''
  Phys.\ Rev.\  D {\bf 49}, 6228 (1994)
  [arXiv:hep-ph/9401301].
  %%CITATION = PHRVA,D49,6228;%%
%\cite{Wang:2006zw}
\bibitem{Wang:2006zw}
  Z.~G.~Wang,
  %``Structure of the axial-vector meson D/s1(2460),''
  J.\ Phys.\ G {\bf 34}, 753 (2007)
  [arXiv:hep-ph/0611271].
  %%CITATION = JPHGB,G34,753;%%
%\cite{Mandelstam:1962mi}
\bibitem{Mandelstam:1962mi}
S.~Mandelstam,
%``Quantum Electrodynamics Without Potentials,''
Annals Phys.\  {\bf 19}, 1 (1962);
%%CITATION = APNYA,19,1;%%
%\cite{Terning:1991yt}
%\bibitem{Terning:1991yt}
J.~Terning,
%``Gauging nonlocal Lagrangians,''
Phys.\ Rev.\ D {\bf 44}, 887 (1991).
%%CITATION = PHRVA,D44,887;%%

\end{thebibliography}
\end{document}